\documentclass[runningheads]{llncs}

\usepackage{accv}

\usepackage{accvabbrv}

\usepackage{graphicx}
\usepackage{booktabs}

\usepackage[accsupp]{axessibility}  % Improves PDF readability for those with disabilities.

\usepackage[dvipsnames]{xcolor}
\usepackage{wrapfig}
\newcommand{\red}[1]{{\color{red}#1}}

\definecolor{tablegreen}{RGB}{11,154,83}
\definecolor{tablered}{RGB}{230,28,100}
\definecolor{tableblue}{RGB}{60,105,225}
\definecolor{tablegray}{gray}{0.45}

\newlength\savedwidth
\newcommand\whline{\noalign{\global\savedwidth\arrayrulewidth\global\arrayrulewidth 0.8pt}
\hline\noalign{\global\arrayrulewidth\savedwidth}}

\usepackage{amsmath,amsfonts,bm}

\def\1{\bm{1}}

\def\rvb{{\mathbf{b}}}

\def\rvs{{\mathbf{s}}}

\def\rvv{{\mathbf{v}}}
\def\rvw{{\mathbf{w}}}
\def\rvx{{\mathbf{x}}}

\def\rmF{{\mathbf{F}}}

\def\rmI{{\mathbf{I}}}

\def\rmK{{\mathbf{K}}}

\def\rmR{{\mathbf{R}}}

\def\rmX{{\mathbf{X}}}

\DeclareMathAlphabet{\mathsfit}{\encodingdefault}{\sfdefault}{m}{sl}
\SetMathAlphabet{\mathsfit}{bold}{\encodingdefault}{\sfdefault}{bx}{n}

\usepackage{multirow}
\usepackage[export]{adjustbox}
\usepackage{svg}
\usepackage{makecell}

\definecolor{tablegreen}{RGB}{245,230,245}
\definecolor{tablered}{RGB}{235,237,245}
\definecolor{tableblue}{RGB}{225,245,245}

\definecolor{Highlight}{HTML}{39b54a}  %
\definecolor{lightgray}{gray}{0.7}%

\usepackage{appendix}
\usepackage{pifont}
\usepackage{xcolor}
\usepackage{colortbl}
\renewcommand{\red}[1]{\textcolor{red}{#1}}

\makeatletter

\newcommand{\Rmnum}[1]{\expandafter\@slowromancap\romannumeral #1@}
\makeatother

\usepackage{arydshln}
\usepackage[breaklinks,colorlinks]{hyperref}
\usepackage{orcidlink}

\begin{document}

\title{PlainUSR: Chasing Faster ConvNet for Efficient Super-Resolution} 

\titlerunning{PlainUSR}

\author{Yan Wang\orcidlink{0000-0002-5989-2408} \and
Yusen Li\orcidlink{0000-0001-6623-350X} \and Gang Wang\orcidlink{0000-0003-0387-2501} \and
Xiaoguang Liu\orcidlink{0000-0002-9010-3278} }
\authorrunning{Wang et al.}

\institute{Nankai University, Tianjin, China 
% \and Springer Heidelberg, Tiergartenstr.~17, 69121 Heidelberg, Germany
\\
\url{https://github.com/icandle/PlainUSR}} 
% \and

\maketitle

\begin{abstract}
Reducing latency is a roaring trend in recent super-resolution (SR) research. While recent progress exploits various \emph{convolutional blocks}, \emph{attention modules}, and \emph{backbones} to unlock the full potentials of the convolutional neural network (ConvNet), achieving real-time performance remains a challenge. To this end, we present PlainUSR, a novel framework incorporating three pertinent modifications to expedite ConvNet for efficient SR. For the convolutional block, we squeeze the lighter but slower MobileNetv3 block into a heavier but faster vanilla convolution by reparameterization tricks to balance memory access and calculations. For the attention module, by modulating input with a regional importance map and gate, we introduce local importance-based attention to realize high-order information interaction within a 1-order attention latency. As to the backbone, we propose a plain U-Net that executes channel-wise discriminate splitting and concatenation. In the experimental phase, PlainUSR exhibits impressively low latency, great scalability, and competitive performance compared to both state-of-the-art latency-oriented and quality-oriented methods. In particular, compared to recent NGswin, the PlainUSR-L is 16.4$\times$ faster with competitive performance. 
 
\end{abstract}

\section{Introduction}
\label{sec:intro}
Image super-resolution (SR) has been a long-consisting low-level vision task aiming at restoring a proper high-resolution (HR) image from its numerous low-resolution degradation. Since the burgeon of deep learning, convolutional SR networks~\cite{FSRCNN,VDSR,EDSR,RCAN} have become mainstream and achieved remarkable success. Despite the rising PSNR, the correspondingly increasing latency limits practical usage on mobile devices (\eg, SR for cloud gaming and live streaming), which forces researchers to refine a faster and more efficient model. Generally, prevailing works developed more efficient ConvNets from three perspectives: 1) \emph{convolution block}, 2) \emph{attention module}, and 3) \emph{backbone}. We revisit these components and ease the ``bottleneck'' restricting their latency reduction.

For \emph{convolution block}, a primary attempt is to adopt residual learning to improve the representation capability and ameliorate optimization. For example, EDSR~\cite{EDSR} utilized residual block~\cite{ResNet} and RFDN~\cite{RFDN} leveraged shallow residual block. While residual learning contributes to heightened restoration quality, it maintains comparatively intensive computations and increased memory usage. To address this limitation, the following works~\cite{VapSR,MAN,BSRN} focused on convolution decomposition to stack numerous simplified convolutions as the replacement for standard convolution. 
A notable example of this decomposition is the MobileNet~\cite{mobilenetv3} block, which decouples convolution into depthwise and pointwise convolutions, showcasing exceptional efficiency across diverse tasks~\cite{mobilenetv3sr}.
Despite remarkably reducing parameters and calculations, the latency is instead growing due to higher memory access, especially for tasks with large input image. For instance, the MBConv is slower than vanilla convolution in the SR task. Inspired by reparameterization technology~\cite{ECB,EFDN} that transfers the complex training-time block into a simple one for inference, we equivalently transform the reparameterizable MBConv (RepMBConv) into a vanilla convolution to reach a more reasonable trade-off. Compared to the original MBConv, RepMBConv has more MACs but lower memory access, attaining lower latency and better extraction.  

For \emph{attention module}, RCAN~\cite{RCAN} first absorbed channel attention into the residual block and achieved prominent performance. Building upon RCAN's success, subsequent works~\cite{HAN,PAN} delved into further explorations of attention mechanisms in super-resolution tasks. Specifically, RFANet~\cite{RFANet} proposed effective spatial attention (ESA), which works its magic on both performance-oriented and lightweight frameworks~\cite{RFDN}. However, these attention mechanisms facilitate only 1-order information interaction with a limited receptive region. Recent studies~\cite{NLRN,SwinIR} apply 2-order interaction, \eg, non-local attention~\cite{NLSA} and self-attention~\cite{ELAN}, to realize more comprehensive long-range modeling. However, these approaches introduce additional runtime due to the quadratic complexity. To strike a balance, we propose local importance-based attention (LIA) that realizes 2-order interaction with extremely simple operators. Specifically, LIA measures the local importance at the downscaled feature map and calibrates the attention map with one channel gate, thereby ensuring a relatively low latency.

For \emph{backbone}, three typical designs are commonly employed: VGG~\cite{RepVGG}-style, ResNet~\cite{ResNet}-style, and IMDN~\cite{IMDN}-style, with the increasing performance but descending throughput. The VGG-style networks, exemplified by ESPCN~\cite{ESPCN} and QuickSRNet~\cite{QuickSRNet},  demonstrate optimal deployment while encountering quality drops when enlarging model size. For prevailing efficient SR models~\cite{RFDN,EFDN,RLFN}, they tend to employ the EDSR/IMDN-like framework for guaranteed performance but ignore latency. In our endeavor to balance performance and latency, we introduce a U-Net design into the plain network by proposing a PlainU-Net, which executes a channel-dimension ``U'' shape processing.

Incorporating the aforementioned refinements: RepConv, LIA, and PlainU-Net, we introduce a new-generation ConvNet termed PlainUSR, aimed at achieving faster runtimes for efficient super-resolution tasks. Our contributions can be summarized as follows:
\begin{enumerate}
    \item[1)] We revisit the efficient SR designs from three fundamental components (convolutional block, attention module, and backbone), and explore and alleviate bottlenecks of latency by following refinements:
    \begin{itemize}
    \item We propose a reparameterized MBConv (RepMBConv) that seamlessly integrates MBConv into vanilla convolution without any performance degradation but 2.9$\times$ acceleration.
    \item We introduce local importance-based attention (LIA), surpassing existing attention mechanisms in terms of both quality and efficiency.
    \item We introduce a novel backbone, PlainU-Net, which remains straightforward during inference yet attains enhanced representation and optimization properties compared to existing backbones.
    \end{itemize}
    \item[2)] Based on these modifications, we present a simple yet effective framework, namely PlainUSR, which records significantly lower latency, greater scalability, and relatively fewer calculations than existing efficient SR models.
\end{enumerate}

\section{Related Work}
\subsection{Efficient Super-Resolution}
To alleviate the complexity of upscaling LR images on devices with constrained resources, numerous strategies~\cite{FSRCNN,IMDN,EFDN} have been introduced for efficient ConvNet design. The representative work, IMDN~\cite{IMDN} employed information multi-distillation and contrast-aware channel attention (CCA), reaching remarkable performance within 1M parameters. Based on IMDN, RFDN~\cite{RFDN} replaced channel splitting operation and CCA with feature distillation connection and more powerful ESA, to improve the flexibility and representation capability. Then, several researches~\cite{RLFN,DIPNet,FMEN} removed the skip connections and utilized structural parameterization to enable lossless improvement on a faster backbone. FMEN~\cite{FMEN} proposed parameterizable residual in residual block (RRRB) and backbone with only one global residual connection, which spent extremely low memory and latency.
Subsequent studies~\cite{ABPN,ETDS} rehashed the plain VGG-like backbone for compact feature processing. However, an unavoidable challenge is the poor optimization of the plain structure. Recent work attempted to resolve the issue by reparameterization. For instance, QuickSRNet~\cite{QuickSRNet} adopted reparameterizable parameter initialization. ETDS~\cite{ETDS} squeezed dual stream network into a plain one. These plain networks obtained better trade-offs between performance and latency.

\subsection{Reparameter and Attention in Efficient Super-Resolution}
\noindent\textbf{Re-parameterization}. Since pioneer RepVGG~\cite{RepVGG} equivalently transformed a three-branch block into a single layer after training for ``free'' improvement, structure parameterization has become an essential strategy for efficient SR.
ECBSR~\cite{ECB} proposed an edge-oriented convolution block (ECB) that leveraged the Soble filter for better edge generating. EFDN~\cite{EFDN} further developed ECB with Laplacian branches and edge loss to achieve better learning capability. SESR~\cite{FMEN} collapsed the linear blocks into single convolution layers to add the training-time network depth for better optimization.

\noindent\textbf{Attention}.
The attention module is an adaptive discriminating selection to retain useful features and automatically ignore noisy responses according to the input.
As to efficient SR, most existing works adopt the existing attention modules.
% For the SR task, RCAN~\cite{RCAN} first utilized channel attention to adaptively rescale each channel-wise feature.
% RFANet~\cite{RFANet} proposed enhanced spatial attention to explore and exploit spatial aggregation. They attain impressive success for performance-oriented tasks.
For instance, RLFN~\cite{RLFN} employed ESA from RFANet~\cite{RFANet} and PAN~\cite{PAN} leveraged pixel-attention~\cite{PiCANet}. Recently, 2-order attention mechanisms, \eg, self-attention~\cite{SwinIR,NGSwin} and non-local attention~\cite{NLSA,ENLCA}, bring higher performance. However, they are rarely used for low-latency SR models due to their computationally expensive costs.

\section{Methodology}

\subsection{Convolution: Reparameterized MBConv}

\begin{figure}[t]
\begin{minipage}[c]{0.62\textwidth}
\includegraphics[width=1\linewidth]{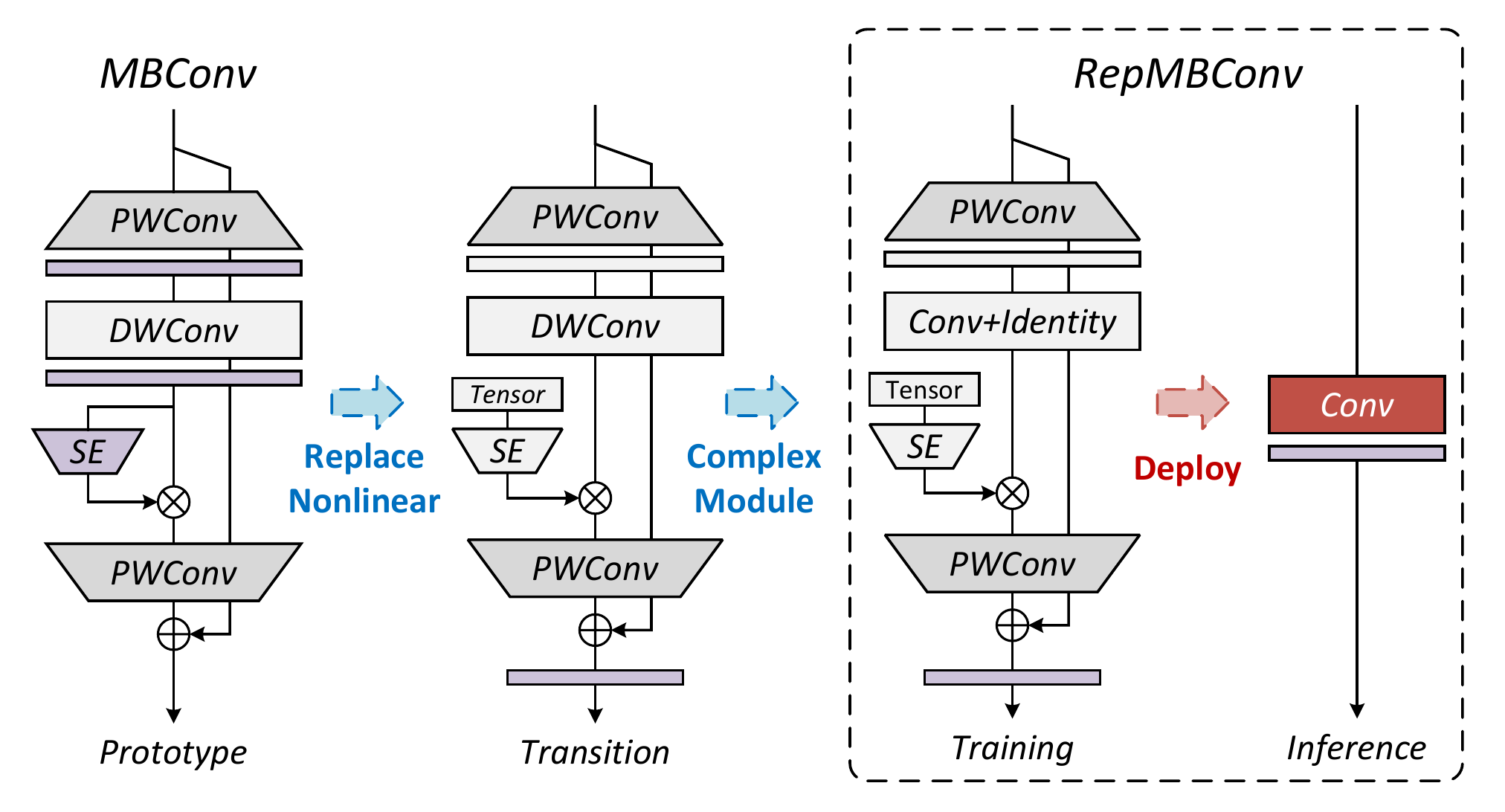}
\captionof{figure}{Illustration of MBConv and RepMBConv.}
\label{fig:RepMBConv}
\end{minipage}
\begin{minipage}[c]{0.36\textwidth}
\vspace{2mm}
\fontsize{8.0pt}{9.5pt}\selectfont
\tabcolsep=1pt
\begin{tabular}{c|cc}
\whline
Metrics & MBConv & Conv \\
\whline
Channel & 64 & 64 \\
Params & {26.0}k  & 36.9k   \\
MACs & {1.01}G & 2.12G  \\
Memory locate &  176.9M & {42.5}M   \\
Memory access &  40.6M & {7.38}M  \\
\hline
% VGGSR & 20.3G & 40.7G \\
Activations & 342.8M & {74.5}M \\
Latency & 110.9ms & {39.3}ms\\
\whline
\end{tabular}
% \vspace{-3mm}
\captionof{table}{Comparisons for MBConv and vanilla convolution with 256$\times$256 inputs. Latency is validated on RTX 4060.}
\label{tab:Operator}
\end{minipage}
\end{figure}

The MobileNetV3~\cite{mobilenetv3} Block (MBConv) has achieved remarkable success in various computer vision tasks~\cite{OmniSR,mobilenetv3sr}. Despite its parameter and calculation efficiency, its higher memory access of depth-wise and point-wise convolution results in non-negligible delay, thereby reducing the running speed. For I/O-bound devices and I/O-demand tasks, this drawback is exacerbated, for example, the efficient super-resolution task that conducts full-size image super-resolution on restricted devices. As indicated in \cref{tab:Operator}, despite MBConv attaining fewer parameters and calculations, the model based on MBConv spends more time than vanilla convolution in $\times$4 SR task on DIV2K-valid~\cite{div2k}.  Drawing inspiration from reparameterization technology~\cite{RepVGG,DBB,ECB}, we aim to address these limitations by reparameterizing the MBConv to a vanilla convolution. In essence, we reparameterize the MBConv to convolution (RepMBConv) through three steps, as illustrated in \cref{fig:RepMBConv}.

\noindent \textbf{Replacing nonlinear}: As the nonlinear operators (GELU~\cite{GELU} activation and SE~\cite{SE} module) are un-reparameterizable, we opt to replace them with the linear operator or move them to the tail. For the activation function, we replace the first GELU with a scaling function and move the second one to the tail. For the SE module, based on \emph{stripe observation}~\cite{ASR} that channel attention's values of varied inputs in the dataset tend towards a constant vector, we replace the input-relative SE with the learnable tensor-relative SE module.

\noindent \textbf{Complex module}: As the RepMBConv is deployed as a vanilla convolution in the inference stage, adding the complexity of training topology is harmless to improve the representation capability. Thus, we replace the depth-wise convolution with standard convolution and identity to chase a better performance as previous works~\cite{DBB,ECB}.

\noindent \textbf{Deployment}: Following~\cite{RepVGG,DBB}, we train the complicated structure and squeeze it to convolution during inference for deployment. Given the input feature $\rmX\in\mathbb{R}^{C\times H\times W}$, the training process of RepMBConv $\mathcal{C}(\cdot)$ can be formulated by: 
\begin{equation}
   \mathcal{C}(\rmX) = \{ \left[\left( \rmX*\textcolor{Mulberry}{\rmK_1}+\textcolor{Mulberry}{\rvb_1}\right)\odot\rvs*\textcolor{Mulberry}{\rmK_2}+\textcolor{Mulberry}{\rvb_2}\right]\odot\mathcal{F}_{SE}(\rvv) *\textcolor{Mulberry}{\rmK_3} + \textcolor{Mulberry}{\rvb_3}\} + \rmX,  
   \label{eq:RepMBConv_train}
\end{equation}
where $\odot$ and $*$ are scale-dot and convolution, respectively. $({\color{Mulberry}{\rmK_1}}\in\mathbb{R}^{C_{m}\times C\times 1\times 1}, {\color{Mulberry}{\rvb_1}}\in\mathbb{R}^{C_{m}})$ are the kernel and bias weight for expanding point-wise convolution, $[{\color{Mulberry}{\rmK_2}}\in\mathbb{R}^{C_{m}\times C_{m}\times k\times k}, {\color{Mulberry}{\rvb_2}}\in\mathbb{R}^{C_{m}}]$ are the reparameterized kernel and bias for immediate $3\times3$ convolution and identity; $\{{\color{Mulberry}{\rmK_3}}\in\mathbb{R}^{C\times C_{m}\times 1\times 1}, {\color{Mulberry}{\rvb_3}}\in\mathbb{R}^{C}\}$ are weights of squeezing point-wise convolution. $\rvs,\rvv\in\mathbb{R}^{C_{m}}$ are learnable scalar and vector to replace nonlinear operators. $\mathcal{F}_{SE}(\cdot)$ is SE module calculating channel attention. By unfolding~\cref{eq:RepMBConv_train}, we can easily obtain the reparameterized kernel and bias for deployment:
\begin{equation}
\begin{aligned}
    \rmK_{Rep} &= {\color{Mulberry}\rmK_1}\odot\rvs*{\color{Mulberry}\rmK_2}\odot\mathcal{F}_{SE}(\rvv)*{\color{Mulberry}\rmK_3}+\rmI, \\
    \rvb_{Rep} &=  ({\color{Mulberry}\rvb_1}\odot\rvs*{\color{Mulberry}\rmK_2} + {\color{Mulberry}\rvb_2})\odot\mathcal{F}_{SE}(\rvv)*{\color{Mulberry}\rmK_3} + {\color{Mulberry}{\rvb_3}}, \\
    \mathcal{C}(\rmX) &= \rmX*\rmK_{Rep} + \rvb_{Rep},
\end{aligned}
\end{equation}
where $\rmK_{Rep}\in\mathbb{R}^{C\times C\times k\times k}$ and $\rvb_{Rep}\in\mathbb{R}^{C}$ are reparameterized weights. $\rmI\in\mathbb{R}^{C\times C\times k\times k}$ is the reparameterized identity operator that satisfies $\rmI[i,i,\lfloor, \frac{k}{2}\rfloor, \frac{k}{2}\rfloor]=1, i\in[0,C)$ but other values are zeros.

\subsection{Attention: Local Importance-based Attention}
\begin{wrapfigure}{r}{5.5cm}
    \centering
    \vspace{-5mm}
    \includegraphics[width=1\linewidth]{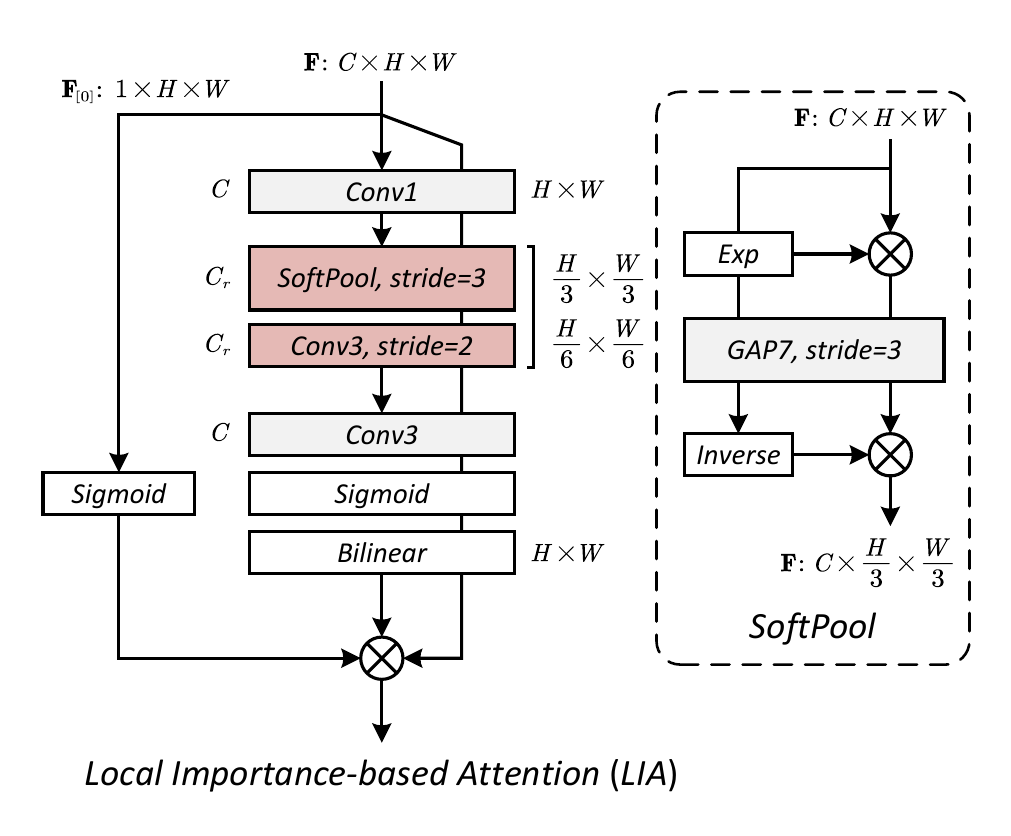}
    \caption{Illustration of the proposed local importance-based attention (LIA).}
    \label{fig:LIP}
    \vspace{-5mm}
\end{wrapfigure}
Another impediment to a faster SR network is the spatial attention calculation. Existing spatial attention mechanisms can be primitively classified according to their orders of interaction~\cite{gnconv}, \eg, ESA~\cite{RFANet} (1-order: one element-wise multiplication) and Self-Attention~\cite{SwinIR} (2-order: two matrix multiplications). However, they suffer different defects that limit their usage in lightweight SR models: the weak performance for 1-order attention and quadric complexity for 2-order attention. 
Thus, we revisit the existing work and tailor a simplified 2-order spatial attention to trade off computation and performance. 

The core of the attention mechanism is to adaptively enhance helpful information but weaken useless one according to their input-relative importance. In prevailing studies, the importance map is measured by subnetwork or matrix multiplication. Inspired by~\cite{LIP,SoftPool} that attains the local importance by regional softmax, we calculate the importance value of the pixel $\rvx$ within the surround $\rmR$ as follows:
\begin{equation}
\mathcal{I}(\rmX)|_\rvx = \sum\limits_{k\in \rmR}\sum\limits_{i\in \rmR_k}{\frac{e^{\rvx_i}}{\sum\limits_{j\in \rmR_k}
e^{\rvx_j}}\cdot\rvw_k}, %\cdot\rvx_i,
\label{eq:importance}
\end{equation}
where $\mathcal{I}(\rmX)|_\rvx$ is the local importance for $\rvx$. $\rmR$ refers to neighborhood centered at $\rvx$. $\rvw$ represents the learnable weight to refine the measured importance. As shown in~\cref{fig:LIP}, we instantiate \cref{eq:importance} by stacking a softpool~\cite{SoftPool} and 3$\times$3 convolution for efficiency and applicability. 
Similar to 1-order attention like ESA, we leverage stride and squeeze convolution to reduce calculation and enlarge the receptive field while sigmoid and bilinear for activation and rescaling.

To re-calibrate the local importance avoiding artifacts brought by stride convolution and bilinear interpolation, we use the gate mechanism~\cite{GLU,MAN} for feature refining of local importance $\mathcal{I(\rmX)}$. Differing from existing gate units applying additional networks, we choose the first channel $\rmX_{[0]}$ maps from the input to perform as a gate for simplicity.
Overall, the LIA $\mathcal{A}(\cdot)$ can be summarised as:
\begin{equation}
\mathcal{A}(\rmX) =  \sigma(\rmX_{[0]}) \odot \psi(\sigma(\mathcal{I(\rmX)})) \odot \rmX,
\label{eq:lia}
\end{equation}
where $\sigma(\cdot)$ and $\psi(\cdot)$ are sigmoid activation and bilinear interpolation.
% Through two element-wise multiplication, we build 2-order attention LIA, which is faster than 1-order ESA and better than 2-order SA. 

\begin{figure}[t]
    \centering
    \includegraphics[width=1.0\linewidth]{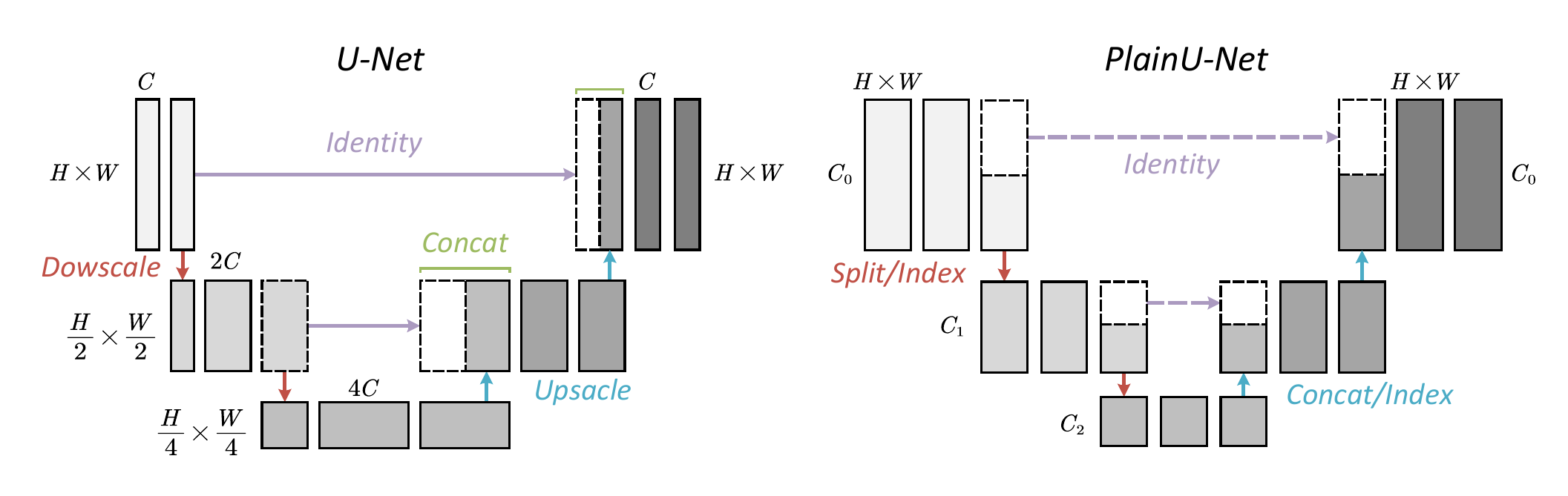}
    \caption{Visual comparison between U-Net and PlainU-Net.}
    \label{fig:PlainU-Net}
\end{figure}

\subsection{Backbone: PlainU-Net}
Despite the above accelerations, the backbone determines the overall latency. Recent studies~\cite{ABPN,QuickSRNet} indicate that a plain backbone (VGG-style) surpasses other backbones, \eg, ResNet~\cite{EDSR} or IMDN~\cite{IMDN} in terms of runtime and memory. Nevertheless, the plain VGG-style backbone uniformly processes the intermediate feature during the forward pass, wasting massive computations. Inspired by U-Net~\cite{UNet}, which effectively encodes and decodes features at the spatial dimension, we introduce the PlainU-Net. This architecture employs a "U" shape processing at the channel dimension. As exhibited in~\cref{fig:PlainU-Net}, the plainU-Net splits and concatenates channel features to conduct hierarchical encoding and decoding throughout the training stage. For inference, it is plain by using a channel index. Given the input feature $\rmF_0 = \rmF \in\mathbb{R}^{C_0\times H\times W}$ and the channel number $C_2\leq C_1\leq C_0$ of each stage, these procedures can be given by:
\begin{equation}
\begin{aligned}
 \rmF_1, \rmF_{0\_} &= \mathcal{S}(\mathcal{H}_1(\rmF_0),[C_1,C_0-C_1]) &\Longleftrightarrow \rmF[:C_0] &= \mathcal{H}_1(\rmF[:C_0]), \\    
 \rmF_2, \rmF_{1\_} &= \mathcal{S}(\mathcal{H}_2(\rmF_1),[C_2,C_1-C_2])  &\Longleftrightarrow  \rmF[:C_1] &= \mathcal{H}_2(\rmF[:C_1]),\\   
 \rmF_3 &= \mathcal{H}_3(\rmF_2)  &\Longleftrightarrow  \rmF[:C_2] &= \mathcal{H}_3(\rmF[:C_2]),\\   
 \rmF_4 &= \mathcal{H}_4(\mathcal{L}(\rmF_3,\rmF_{1\_}))  &\Longleftrightarrow \rmF[:C_1] &= \mathcal{H}_4(\rmF[:C_1]),\\
\rmF_5 &= \mathcal{H}_5(\mathcal{L}(\rmF_4,\rmF_{0\_}))  &\Longleftrightarrow \rmF[:C_0] &= \mathcal{H}_5(\rmF[:C_0]),\\
\end{aligned}
\label{eq:PlainUNet}
\end{equation}
where $\mathcal{H}_i(\cdot)$, $\mathcal{S}(\cdot)$, and $\mathcal{L}(\cdot)$ are convolutional block, split function, and concatenation function, respectively. The left functions formulate the training process while the right ones indicate inference deployment, which employs non-residual forward and shares the memory space of $\rmF$. 

\begin{figure}[t]
    \centering
    \includegraphics[width=0.8\linewidth]{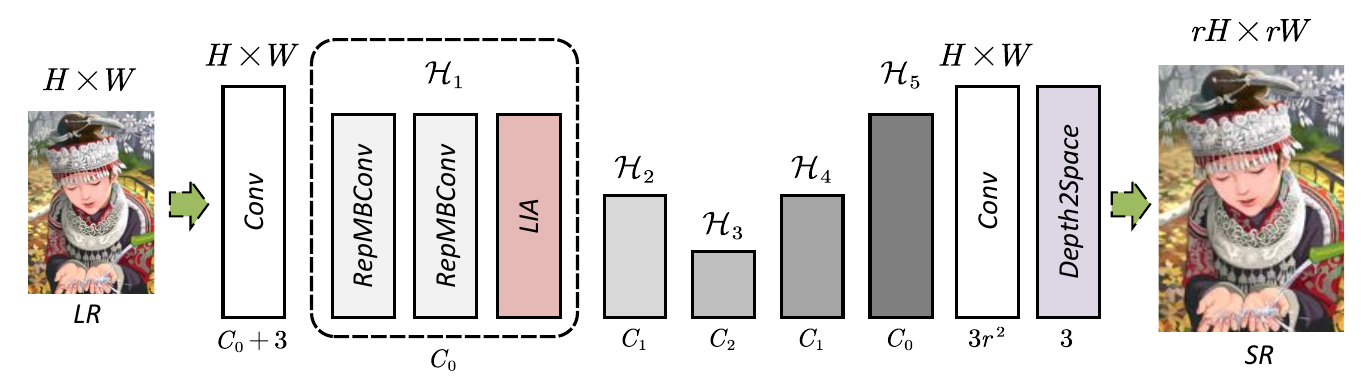}
    \caption{Overview of the proposed PlainUSR framework.}
    \label{fig:PlainUSR}
\end{figure}

\subsection{Overall: PlainUSR}
Building upon RepMBConv, LIA, and PlainU-Net, we build a framework named PlainUSR. As depicted in~\cref{fig:PlainUSR}, we stack two RepMBConv and an LIA to instantiate $\mathcal{H}_i$ in \cref{eq:PlainUNet}. In addition to the PlainU-Net backbone, we incorporate a shallow extraction head and reconstruction tail, consistent with prior work~\cite{RLFN}.

\section{Experiment}
\subsection{Experimental setup}
\noindent \textbf{Datasets and metrics}. Following prevailing works~\cite{li2022ntire,RFDN,EFDN}, we utilize DF2K (800 images from DIV2K~\cite{div2k} and 2650 images from Flickr2K~\cite{EDSR}) as the training datasets. During the testing phase, we assess our models on five classic benchmarks, including Set5~\cite{Set5}, Set14~\cite{Set14}, BSD100~\cite{B100}, Urban100~\cite{Urban100}, and DIV2K-valid~\cite{div2k}. Additionally, following~\cite{ClassSR,ARM}, we utilize DIV8K~\cite{div8k} to generate Test2K and Test4K as more comprehensive benchmarks to prevent model overfitting on traditional test sets. For quality evaluation, we use PSNR and SSIM~\cite{SSIM} metrics. For efficiency evaluation, referring to NTIRE challenge~\cite{li2022ntire}, we exploit parameters, FLOPs, latency, activations, and max GPU memory footprint to examine the proposed models with state-of-the-art methods.

\noindent\textbf{Training details}. 
We implement PlainUSR-\textbf{U}tra/\textbf{T}iny/\textbf{S}mall/\textbf{M}edium/\textbf{B}ase/ \textbf{L}arge with varied settings to explore comprehensive trade-offs between performance and complexity. Specifically, for \textbf{U}ltra/\textbf{T}iny models, we leverage 2-stage PlainU-Net backbones with channel numbers \{16, 8\} and \{32, 16\}. For \textbf{S}mall/\textbf{M}edium/\textbf{B}ase/\textbf{L}arge models, we employ 3-stage PlainU-Net and set channel numbers \{32, 16, 8\}, \{48, 32, 16\}, \{64, 48, 32\}, and \{80, 64, 48\}, respectively. Regarding the training procedure, we randomly sample 64$\times$64 LR images with a batch size of 64 for each iteration. The optimization objective is the $\ell_1$ loss, and we employ AdamW~\cite{AdamW}($\beta_1$=0.9, $\beta_2$=0.99) to optimize PlainUSR. The learning rate is initialized as 5$\times$10$^{-4}$ and halved at \{250\emph{k}, 400\emph{k}, 450\emph{k}, 475\emph{k}\} within 500\emph{k} iterations. All experiments are conducted using the PyTorch~\cite{Pytorch} framework on 4 Tesla V100 GPUs.

\begin{table*}[!t]
\center
\small
\footnotesize
%\scriptsize
\fontsize{7.5pt}{9.5pt}\selectfont
\tabcolsep=1pt
\begin{center}
\caption{Quantitative comparison (average PSNR/SSIM on RGB, Parameters, MACs, Latency, Memory Footprint, and Activations) with state-of-the-art approaches for efficient image SR ($\times$4).
Overall best results are in \red{red}. MACs are measured under the setting of the input image to 256$\times$256. Latency and Memory footprint are reported on the DIV2K-valid~\cite{div2k} dataset with Mobile AMD Ryzen 7940hs (CPU), RTX 4060-Laptop (GPU1) and Tesla V100 (GPU2).} 

\label{tab:lightweight_sr_results}
  % Here 1/2

\begin{tabular}{l|ccc|cccc|c|c|c}
\whline
\multirow{2}{*}{Method} & {CPU} & {GPU1} & {GPU2} & {Para} & {MACs} & {Mem} & Acts & {DIV2KV~\cite{div2k}} &  {Test2K~\cite{div8k}} & {Test4K~\cite{div8k}}
\\
% \cline{5-9}
& (ms) & (ms) & (ms) & (K) & (G) & (M) & (M) & PSNR/SSIM & PSNR/SSIM & PSNR/SSIM 
\\
\whline
IMDN~\cite{IMDN} & 692 & 78.3 & 32.6 %& 
& 893 & 58.53 & 473.7 & 154.14
& \red{29.13}/0.8221 %30.59/0.8409
% & 26.29/0.7141 %27.63/0.7381
& 26.24/0.7614 %27.65/0.7762
& 27.71/0.8078
\\
RFDN~\cite{RFDN} & 680 & 56.5 & 25.8 %& 13.8 
& 433 & 27.10 & 832.0 & 112.03
& 29.04/0.8201 %30.51/0.8389
% & 26.26/0.7130 %27.61/0.7369
& 26.23/0.7601 %27.63/0.7747
% & 30.59/0.9086
& 27.70/0.8067
\\
\rowcolor{tablered}
PlainUSR-L & {554} & {49.8} & {23.0} %& 8.9
& 734 & 44.22 & {327.7} & {80.62}  
& 29.12/\red{0.8221} %30.42/0.8373
% & 26.20/0.7111 %27.54/0.7351
& \red{26.29/0.7622} %27.59/0.7728
% & 30.24/0.9038
& \red{27.77/0.8086}
\\
% PCEVA~\cite{PCEVA} & 468 & 44.7 & 20.9 %& 9.9
% & 402 & 24.73 & 372.5 & 72.09
% & 28.68/0.8104 %30.14/0.8300
% % & 26.31/0.7123 %27.66/0.7364
% & 26.01/0.7511 %27.42/0.7664
% % & 29.81/0.8993
% & 27.41/0.7981
% \\
EFDN~\cite{EFDN} & 420 & 43.4 & 20.4  %& 13.8 
& {276} & {16.73} & 710.0 & 111.12  
& 29.00/0.8187 %30.51/0.8389
% & 26.21/0.7110 %27.61/0.7369
& 26.17/0.7580 %27.63/0.7747
% & 30.59/0.9086
& 27.63/0.8049
\\
ECBSR~\cite{ECB} & 413 & 39.7 & 19.7 %& 9.5
& 622 & 40.66 & {231.6}  & 77.59
& 28.86/0.8190 %30.46/0.8378
% & 26.22/0.7114 %27.57/0.7359
& 26.13/0.7564
% & 30.49/0.9073
& 27.57/0.8033
\\
FMEN~\cite{FMEN} & 629 & 31.5 & 18.2 %& 9.5
& 341 & 22.28 & \red{205.9} & {72.09}
& 29.00/0.8190 %30.46/0.8378
% & 26.22/0.7114 %27.57/0.7359
& 26.17/0.7586
% & 30.49/0.9073
& 27.62/0.8051
\\
RLFN~\cite{RLFN} & 346 & 33.9 & 17.1 %& 9.5
& 317 & 19.70 & 521.5 & 80.05
& 29.00/0.8190 %30.46/0.8378
% & 26.23/0.7120 %27.57/0.7359
& 26.20/0.7593
& 27.67/0.8060
% & 30.49/0.9073
\\
% QuickSRNet~\cite{QuickSRNet} & \blue{294} & 29.2 & \blue{15.3}  %& 9.5
% & 436 & 28.51 & 296.5 & \blue{56.62}
% & 28.78/0.8140 %30.46/0.8378
% % & 26.23/0.7120 %27.57/0.7359
% & 26.08/0.7552
% & 27.49/0.8019
% % & 30.49/0.9073
% \\
% PFDN~\cite{li2023ntire} & 298 & 31.0 & 16.3 %& 6.3
% & \blue{272} & 16.76 & 344.3 & \blue{65.10}
% & 28.95/0.8176 %30.42/0.8368
% % & 27.54/0.7349
% & 26.17/0.7578
% & 27.63/0.8047
% % & 30.33/0.9051
% \\
DIPNet~\cite{DIPNet} & {343} & {29.1} & {16.0} %& 9.9
& \red{243} & \red{14.90} & 550.4 & 72.97
& 29.04/0.8184 %30.51/0.8379
% & 26.11/0.7074 %27.45/0.7318
& 26.11/0.7550 %27.51/0.7700
% & 29.81/0.8993
& 27.52/0.8018 
\\
\rowcolor{tablered}
PlainUSR-B & \red{289} & \red{26.8} & \red{14.1} %& 8.9
& 333 & 18.69 & {327.7} & \red{46.93}  
& 28.96/0.8181 %30.42/0.8373
% & 26.20/0.7111 %27.54/0.7351
& 26.18/0.7579 %27.59/0.7728
% & 30.24/0.9038
& 27.63/0.8048
\\
%\hline

\whline 
\end{tabular}
\end{center}
\end{table*}

\subsection{Comparison with SOTA SR models}
To demonstrate the performance of the proposed PlainUSR framework, we compare varied PlainUSR with the state-of-the-art models.

\begin{table}[t]
% \begin{table}[t]
\centering
\caption{Quantitative comparison (average PSNR/SSIM on Y of YCbCr, Parameters, MACs, Latency, and Activation) with state-of-the-art approaches for varied trade-offs in image SR ($\times$4).
The models are divided into six groups according to their throughput on the Tesla V100, \ie, from top to bottom: \emph{240 fps} ($<$4.16ms), \emph{180 fps} ($<$5.55ms), \emph{120 fps} ($<$8.33ms), \emph{90 fps} ($<$11.11ms), \emph{60 fps} ($<$16.66ms), and others.
The best results of each group are in \red{red}. For this and the following tables~\cref{tab:srx2,tab:sr3}, MACs are measured under the setting of the input image to 256$\times$256. Latency is reported on the DIV2K-valid~\cite{div2k} dataset with (i) RTX 4060-Laptop (ms) and (ii) Tesla V100 (ms).} 
% \fontsize{6.6pt}{9.5pt}\selectfont
\tabcolsep=0.5pt
\fontsize{6.6pt}{9.5pt}\selectfont
\begin{tabular}{l|cc|ccc|c|c|c|c}
\whline
\multirow{2}{*}{Method ($\times$4)}  & \multicolumn{2}{c|}{Latency}&{Para} & {MACs} & {Acts} & Set5~\cite{Set5}  & Set14~\cite{Set14} & BSD100~\cite{B100} & Urban\cite{Urban100}\\
& (i) & (ii) & (K) & (G) & (M) & PSNR/SSIM & PSNR/SSIM & PSNR/SSIM & PSNR/SSIM  \\
\whline
ESPCN~\cite{ESPCN} & {10.9} & \red{2.7} & 37.2 & 2.43 & {9.44} 
& 30.66/0.8688
& 27.66/0.7581 
& 26.94/0.7152
& 24.56/0.7263
\\
% ETDS-T~\cite{ETDS} &  & {11.1} & {3.7} & 21.4 & 1.39 & {12.58} 
% & 30.87/0.8738
% & 27.75/0.7618 
% & 27.04/0.7194
% & 24.71/0.7350
% \\
\rowcolor{tablered}
PlainUSR-U & \red{9.5} & {3.2} & {19.8} & {1.29} & {10.16} 
& \red{30.77/0.8698}
& \red{27.72/0.7602}
& \red{27.03/0.7173}
& \red{24.71/0.7327}
\\
\whline
% \hline
ECBSR-{\tiny{M4C16}}\cite{ECB} & {10.4} & {5.5} & 16.8 & 1.09 & {8.39} 
& 30.89/0.8735
& 27.81/0.7623
& 27.06/0.7181
& 24.76/0.7366
\\
QuickSRNet-S~\cite{QuickSRNet}& \red{10.1} & {4.6} & 33.3 & 2.17 & {9.44} 
& 30.91/0.8746
& 27.85/0.7627
& 27.06/0.7183
& 24.76/0.7373
\\
ETDS-S~\cite{ETDS} & {11.6} & {4.7} & 54.1 & 3.53 & {16.78} 
& 31.19/0.8806 
& 28.01/0.7678 
& 27.18/{0.7226}
& 25.03/0.7479
\\
\rowcolor{tablered}
PlainUSR-T & {10.8} & \red{4.5} & 60.1 & 3.93 & {16.97} 
& \red{31.23/0.8809} 
& \red{28.03/0.7678}
& \red{27.20/0.7230}
& \red{25.04/0.7480}
\\
\whline
% \hline
ECBSR-{\tiny{M4C32}}~\cite{ECB} & {12.5} & {7.8} & {51.9} & {3.39} & {13.63} 
& 31.26/0.8809
& 28.06/0.7683
& 27.22/0.7234
& 25.08/0.7496
\\
QuickSRNet-M~\cite{QuickSRNet} & \red{12.1} & {6.2} & 61.0 & 3.98 & {15.73} 
& 31.35/0.8827
& 28.10/0.7690
& 27.22/0.7241
& 25.10/0.7508
\\
ETDS-M~\cite{ETDS}  & {12.1} & \red{5.6} & 72.6 & 4.74 & {20.97} 
& 31.41/0.8843
& 28.13/0.7705
& 27.27/0.7251
& 25.20/0.7544
\\
\rowcolor{tablered}
PlainUSR-S & {12.4} & {7.1} & 68.6 & 4.19 & {20.60} 
& \red{31.56/0.8855}
& \red{28.17/0.7714}
& \red{27.30/0.7268}
& \red{25.32/0.7597}
\\
\whline
% \hline
ECBSR-{\tiny{M6C40}}~\cite{ECB}& \red{15.8} & {8.7} & 105.4 & 6.88 & {21.50} 
& 31.42/0.8825
& 28.21/0.7714
& 27.33/0.7265
& 25.38/0.7602
\\
PCEVA-M~\cite{PCEVA}  & {16.4} & {10.3} & {67.7} & {4.32} & {19.16} 
& 31.58/0.8867
& 28.25/0.7718  
& 27.36/0.7279
& 25.48/0.7618
\\
ETDS-L~\cite{ETDS} & {16.8} & \red{8.5} & 170.0 & 11.11 & {31.46} 
& 31.69/0.8889
& 28.31/0.7751 
& 27.37/0.7302
& 25.47/0.7643
\\
\rowcolor{tablered}
PlainUSR-M & {18.1} & {10.4} & 165.5 & 9.82 & {32.56} 
& \red{31.79/0.8889}
& \red{28.38/0.7760}
& \red{27.43/0.7313}
& \red{25.68/0.7716}
\\
\whline
% \hline
DIPNet~\cite{DIPNet} & {29.1} & {16.0} & {243.3} & {14.90} & {72.97} 
& 31.81/0.8898
& 28.44/0.7773
& 27.45/0.7318
& 25.75/0.7751
\\
ECBSR-{\tiny{M16C64}}~\cite{ECB} & {29.9} & {15.3} & 621.6 & 40.66 & {77.59} 
& 31.92/0.8946
& 28.34/0.7798
& 27.48/0.7339
& 25.81/0.7773
\\
QuickSRNet-L~\cite{QuickSRNet} & {29.3} & {15.3} & 435.9 & 28.51 & {56.62} 
& 31.88/0.8913
& 28.46/0.7788
& 27.48/0.7332
& 25.76/0.7754
\\
\rowcolor{tablered}
PlainUSR-B  & \red{26.8} & \red{14.1} & 333.1 & 18.69 & {46.93} 
& \red{32.02/0.8922}
& \red{28.54/0.7800}
& \red{27.54/0.7351}
& \red{25.97/0.7814}
\\
\whline
% \hline
IMDN~\cite{IMDN} & 78.3 & 32.6 & {893.9} & {58.53} & {154.1} 
& 32.21/0.8948
& 28.58/0.7811
& 27.56/0.7353
& 26.04/0.7838
\\
RLFN~\cite{RLFN} & {62.7} & {24.7} & 543.7 & 33.99 & {126.5} 
& 32.24/0.8952
& 28.62/0.7813
& 27.60/0.7364
& 26.17/0.7877
\\
DDistill-SR~\cite{DDistillSR} & {130.2} & {50.7} & 675.8 & 37.30 & {226.6} 
& 32.28/\red{0.8961}
& 28.69/0.7833
& 27.64/0.7383
& 26.25/0.7892
\\
\rowcolor{tablered}
PlainUSR-L  & \red{49.8} & \red{23.0} & 734.0 & 44.22 & {80.62} 
& \red{32.28}/0.8959
& \red{28.70/0.7838}
& \red{27.65/0.7385}
& \red{26.35/0.7943}
\\
\whline
\end{tabular}
\label{tab:comp_sr_x4}
\vspace{-6mm}
\end{table}

\begin{figure}[!t]
\centering
\begin{tabular}{cc}
\\
\includegraphics[width=0.49\linewidth]{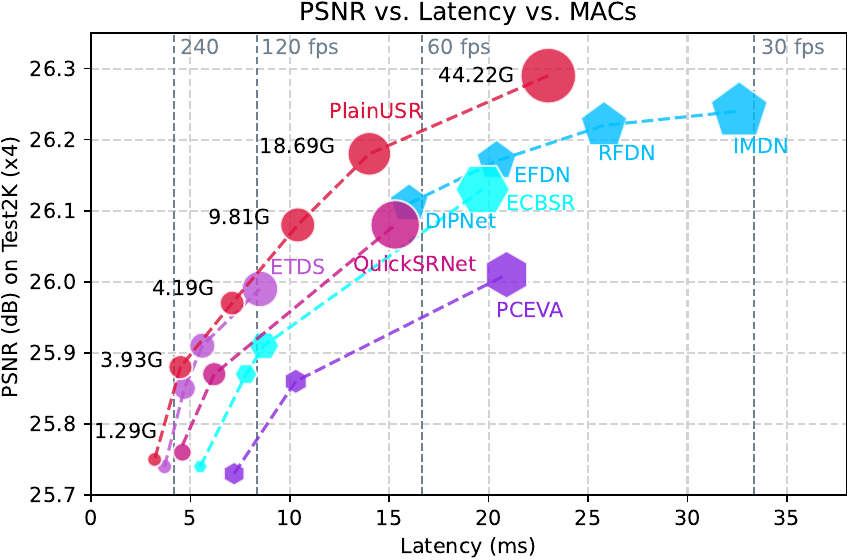}  & 
\includegraphics[width=0.49\linewidth]{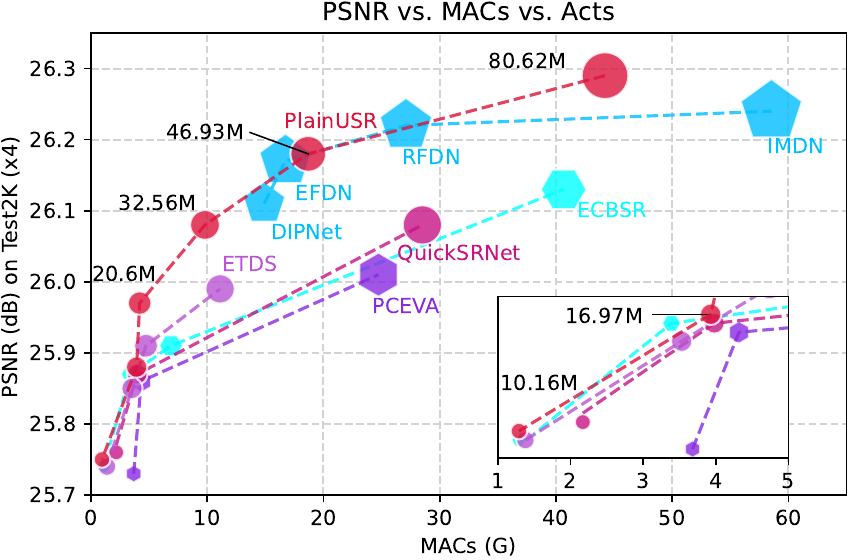} \\
\end{tabular}
\caption{Trade-off curves between restoration performance and runtime complexity.}
\label{fig:tradeoffs}
\end{figure}

\noindent\textbf{Quantitative comparison under similar performance}. Following NTIRE series Efficient SR competitions~\cite{li2022ntire,li2023ntire}, we compare the models under similar performance (around 29.00dB on DIV2K-valid~\cite{div2k}, including IMDN~\cite{IMDN}, RFDN~\cite{RFDN}, EFDN~\cite{EFDN}, ECBSR~\cite{ECB}, FMEN~\cite{FMEN}, RLFN~\cite{RLFN}, DIPNet~\cite{DIPNet}, and the PlainUSR-B/L). To avoid overfitting on the specific dataset, we also add Test2K and Test4K for fair performance validation. As shown in \cref{tab:lightweight_sr_results}, our PlainUSR-B attains the lowest latency and fewest activations while PlainUSR-L reaches the highest PSNR/SSIM. Specifically, compared to DIPNet~\cite{DIPNet}, PlainUSR-B obtains {0.11}dB improvement on Test4K and 10\%-15\%faster. Compared to more complex IMDN~\cite{IMDN}, PlainUSR-L advances by 0.06dB but is 30\% faster.

\noindent\textbf{Quantitative comparison under similar latency}. We also execute comparisons under different performance-complexity trade-offs to demonstrate the scalability and generality of the proposed framework. \cref{tab:comp_sr_x4} exhibits the $\times$4 SR results, where our PlainUSR family achieves incredible improvements over prevailing models, additionally including ESPCN~\cite{ESPCN}, QuickSRNet~\cite{QuickSRNet}, PCEVA~\cite{PCEVA}, ETDS~\cite{ETDS}, and DDistill-SR~\cite{DDistillSR}. In general, the PlainUSR can maintain a comparative throughput (fps) while advancing 0.1dB on Urban100 for almost all groups. Particularly, the PlainUSR-B/M is as fast as QuickSRNet-L/PCEVA-M but attains 0.21dB improvement on Urban100. Compared to ESPCN, PlainUSR-U accelerates 1.4ms and leads by 0.15dB. To enable a more comprehensive comparison, we validate these models on Test2K~\cite{div8k} and visualize them in \cref{fig:tradeoffs}. Specifically, the latency-PSNR curve shows the PlainUSR is more competitive than other plain architecture, \eg, ETDS~\cite{ETDS} and QuickSRNet~\cite{QuickSRNet}. Moreover, compared to these latency-oriented models, our PlainUSR can reach better trade-offs between computations and performance. We also provide $\times$2 results in \cref{tab:srx2}, where PlainUSR consistently improves performance in terms of both PSNR/SSIM under similar latency. These upshots demonstrate the effectiveness of the PlainUSR framework.

\begin{table}[t]
\centering
\caption{Quantitative comparison (average PSNR/SSIM on Y of YCbCr, Parameters, MACs, Latency, and Activation) with state-of-the-art approaches for varied trade-offs in image SR ($\times$2). The best results of each group are in \red{red}.
% The models are divided into six groups according to their throughput on the Tesla V100, \ie, from top to bottom: \emph{240 fps} ($<$4.16ms), \emph{180 fps} ($<$5.55ms), \emph{120 fps} ($<$8.33ms), \emph{90 fps} ($<$11.11ms), \emph{60 fps} ($<$16.66ms), and others.
% The best results of each group are in \red{red}. MACs are measured under the setting of the input image to 256$\times$256. Latency is reported on the DIV2K-valid~\cite{div2k} dataset with (i) RTX 4060-Laptop (ms) and (ii) Tesla V100 (ms).
} 
\fontsize{6.7pt}{9.5pt}\selectfont
\tabcolsep=0.6pt
\begin{tabular}{l|cc|ccc|c|c|c|c}
\whline
\multirow{2}{*}{Method ($\times$2)}  & \multicolumn{2}{c|}{Latency}&{Para} & {MACs} & {Acts} & Set5~\cite{Set5}  & Set14~\cite{Set14} & BSD100~\cite{B100} & Urban~\cite{Urban100}\\
& (i) & (ii) & (K) & (G) & (M) & PSNR/SSIM & PSNR/SSIM & PSNR/SSIM & PSNR/SSIM  \\
\whline
ESPCN~\cite{ESPCN} & {19.2} & {4.9} & 26.8 & 1.75 & {7.08} 
& 36.87/0.9559
& 32.62/0.9086 
& 31.40/0.8898
& 29.61/0.8973
\\
% ETDS-T~\cite{ETDS} &  & {11.1} & {3.7} & 21.4 & 1.39 & {12.58} 
% & 30.87/0.8738
% & 27.75/0.7618 
% & 27.04/0.7194
% & 24.71/0.7350
% \\
\rowcolor{tablered}
PlainUSR-U & \red{17.4} & \red{4.4} & {13.6} & {0.89} & {7.80} 
& \red{37.10/0.9569}
& \red{32.75/0.9101}
& \red{31.56/0.8914}
& \red{29.98/0.9037}
\\
\whline
% \hline
ECBSR-{\tiny{M4C16}}\cite{ECB} & {21.2} & {5.9} & 11.5 & 0.75 & {6.03} 
& 37.18/0.9577
& 32.79/0.9105
& 31.61/0.8928
& 30.18/0.9064
\\
QuickSRNet-S~\cite{QuickSRNet}& \red{20.6} & \red{5.3} & 22.9 & 1.49 & {7.08} 
& 37.18/0.9573
& 32.82/0.9104
& 31.61/0.8921
& 30.17/0.9064
\\
ETDS-S~\cite{ETDS} & {23.8} & {6.8} & 41.5 & 2.71 & {12.06} 
& 37.38/\red{0.9587}
& 32.96/0.9124 
& 31.77/0.8943
& \red{30.62/0.9121}
\\
\rowcolor{tablered}
PlainUSR-T & {21.4} & {6.5} & 48.7 & 3.18 & {14.61} 
& \red{37.47}/0.9584 
& \red{32.97/0.9124}
& \red{31.77/0.8943}
& {30.53/0.9110}
\\
\whline
% \hline
ECBSR-{\tiny{M4C32}}~\cite{ECB} & \red{24.4} & \red{7.2} & {41.5} & {2.71} & {11.27} 
& 37.39/0.9587 
& 33.03/0.9126
& 31.79/0.8953 
& 30.69/0.9128
\\
QuickSRNet-M~\cite{QuickSRNet} & {25.5} & {7.5} & 50.6 & 3.30 & {13.37} 
& 37.42/0.9584
& 33.05/0.9127
& 31.81/0.8951
& 30.77/0.9140
\\
ETDS-M~\cite{ETDS}  & {29.8} & {8.9} & 60.0 & 3.92 & {16.25} 
& 37.54/0.9593
& 33.09/0.9133
& 31.86/0.8963
& 30.87/0.9149
\\
\rowcolor{tablered}
PlainUSR-S & {37.3} & {10.1} & 57.2 & 3.45 & {18.24} 
& \red{37.64/0.9593}
& \red{33.16/0.9146}
& \red{31.93/0.8968}
& \red{31.06/0.9178}
\\
\whline
% \hline
ECBSR-{\tiny{M6C40}}~\cite{ECB}& \red{49.5} & \red{14.2} & 92.4 & 6.03 & {19.14} 
& 37.61/0.9596
& 33.18/0.9139
& 31.94/0.8972
& 31.09/0.9174
\\
% PCEVA-M~\cite{PCEVA}  & {16.4} & {10.3} & {67.7} & {4.32} & {19.16} 
% & 37.83/0.9603
% & 28.25/0.7718  
% & 27.36/0.7279
% & 25.48/0.7618
% \\
ETDS-L~\cite{ETDS} & {53.8} & {17.8} & 152.2 & 9.95 & {26.74} 
& 37.64/0.9597
& 33.24/0.9145 
& 31.98/0.8977
& 31.22/0.9188
\\
\rowcolor{tablered}
PlainUSR-M & {53.2} & {16.7} & 148.9 & 8.74 & {30.20} 
& \red{37.79/0.9597}
& \red{33.37/0.9166}
& \red{32.06/0.8984}
& \red{31.53/0.9228}
\\
\whline
% \hline
% DIPNet~\cite{DIPNet} & {29.1} & {16.0} & {243.3} & {14.90} & {72.97} 
% & 31.81/0.8898
% & 28.44/0.7773
% & 27.45/0.7318
% & 25.75/0.7751
% \\
ECBSR-{\tiny{M16C64}}~\cite{ECB} & {166.9} & {50.0} & 600.7 & 39.29 & {72.88} 
& 37.90/0.9615 
& 33.34/0.9178
& 32.10/0.9018
& 31.71/0.9250
\\
QuickSRNet-L~\cite{QuickSRNet} & {121.6} & \red{34.2} & 415.0 & 27.14 & {51.90} 
& 37.87/0.9600
& 33.45/0.9164
& 32.10/0.8988
& 31.76/0.9246
\\
\rowcolor{tablered}
PlainUSR-B  & \red{117.2} & {35.2} & 311.4 & 17.27 & {44.57} 
& \red{37.95/0.9603}
& \red{33.60/0.9184}
& \red{32.16/0.8996}
& \red{32.00/0.9273}
\\
\whline
% \hline
IMDN~\cite{IMDN} & 338.0 & 96.2 & {873.2} & {57.17} & {151.8} 
& 38.00/0.9605
& 33.63/0.9177
& 32.19/0.8996
& 32.17/0.9283
\\
RLFN~\cite{RLFN} & {232.7} & {75.7} & 526.9 & 32.89 & {124.1} 
& 38.07/0.9607
& 33.72/0.9187
& 32.22/0.9000
& 32.33/0.9299
\\
% DDistill-SR~\cite{DDistillSR} & {130.2} & {50.7} & 675.8 & 37.30 & {226.6} 
% & 32.28/\red{0.8961}
% & 28.69/0.7833
% & 27.64/0.7383
% & 26.25/0.7892
% \\
\rowcolor{tablered}
PlainUSR-L  & \red{219.1} & \red{67.4} & 707.1 & 42.46 & {78.29} 
& \red{38.07/0.9607}
& \red{33.82/0.9202}
& \red{32.27/0.9009}
& \red{32.53/0.9320}
\\
\whline
\end{tabular}
\label{tab:srx2}
\end{table}

\begin{figure*}[!t]
\tabcolsep=1.0pt
\centering
\fontsize{6.0pt}{9.5pt}\selectfont
\renewcommand\arraystretch{0.8}
\begin{tabular}{ccccccccccccc}
\multirow{-5.3}{*}
{\includegraphics[width=.14\linewidth, height=.25\linewidth]{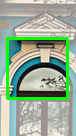}} 
& \includegraphics[width=.11\linewidth, height=.11\linewidth]{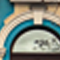}
& \includegraphics[width=.11\linewidth, height=.11\linewidth]{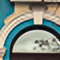}
& \includegraphics[width=.11\linewidth, height=.11\linewidth]{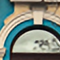}
&
    
&\multirow{-5.3}{*} {\includegraphics[width=.14\linewidth, height=.25\linewidth]{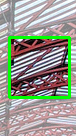}} 
& \includegraphics[width=.11\linewidth, height=.11\linewidth]{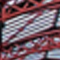} 
& \includegraphics[width=.11\linewidth, height=.11\linewidth]{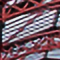}
& \includegraphics[width=.11\linewidth, height=.11\linewidth]{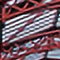}
\\
& Bicubic & ESPCN & PlainUSR-U 
& \quad 
&   & Bicubic & ESPCN & PlainUSR-U  
\\
& {\includegraphics[width=.11\linewidth, height=.11\linewidth]{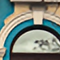}} 
& \includegraphics[width=.11\linewidth, height=.11\linewidth]{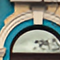}
& \includegraphics[width=.11\linewidth, height=.11\linewidth]{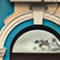}
&   &
& {\includegraphics[width=.11\linewidth, height=.11\linewidth]{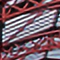}} 
& \includegraphics[width=.11\linewidth, height=.11\linewidth]{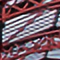}
& \includegraphics[width=.11\linewidth, height=.11\linewidth]{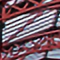}
\\
{\emph{img\_088}~\cite{Urban100}}  
& ECBSR-{\tiny{C16}} & QuickSR-S & PlainUSR-T 
&  
& {\emph{img\_098}~\cite{Urban100}}
& ECBSR-{\tiny{C16}} & QuickSR-S & PlainUSR-T 
\\
\multirow{-5.3}{*}
{\includegraphics[width=.14\linewidth, height=.25\linewidth]{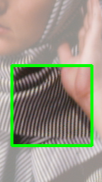}} 
& \includegraphics[width=.11\linewidth, height=.11\linewidth]{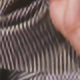}
& \includegraphics[width=.11\linewidth, height=.11\linewidth]{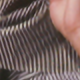}
& \includegraphics[width=.11\linewidth, height=.11\linewidth]{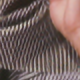}
&
    
&\multirow{-5.3}{*} {\includegraphics[width=.14\linewidth, height=.25\linewidth]{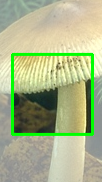}} 
& \includegraphics[width=.11\linewidth, height=.11\linewidth]{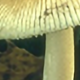} 
& \includegraphics[width=.11\linewidth, height=.11\linewidth]{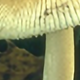}
& \includegraphics[width=.11\linewidth, height=.11\linewidth]{Figure/208001MS/208001_LRBI_x2_PlainUSR_S_x2.png}
\\
& QuickSR-M  & ETDS-M & PlainUSR-S 
& \quad 
&   & QuickSR-M & ETDS-M & PlainUSR-S 
\\
& {\includegraphics[width=.11\linewidth, height=.11\linewidth]{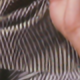}} 
& \includegraphics[width=.11\linewidth, height=.11\linewidth]{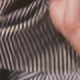}
& \includegraphics[width=.11\linewidth, height=.11\linewidth]{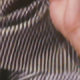}
&   &
& {\includegraphics[width=.11\linewidth, height=.11\linewidth]{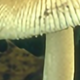}} 
& \includegraphics[width=.11\linewidth, height=.11\linewidth]{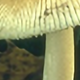}
& \includegraphics[width=.11\linewidth, height=.11\linewidth]{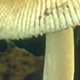}
\\
{\emph{barbara}~\cite{Set14}}  
& QuickSR-L & ETDS-L & PlainUSR-M 
&  
& {\emph{208001}~\cite{B100}}
& QuickSR-L & ETDS-L & PlainUSR-M  
\\
\multirow{-5.3}{*}
{\includegraphics[width=.14\linewidth, height=.25\linewidth]{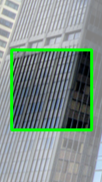}} 
& \includegraphics[width=.11\linewidth, height=.11\linewidth]{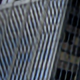}
& \includegraphics[width=.11\linewidth, height=.11\linewidth]{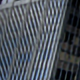}
& \includegraphics[width=.11\linewidth, height=.11\linewidth]{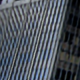}
&
    
&\multirow{-5.3}{*} {\includegraphics[width=.14\linewidth, height=.25\linewidth]{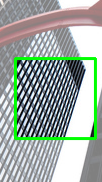}} 
& \includegraphics[width=.11\linewidth, height=.11\linewidth]{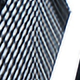} 
& \includegraphics[width=.11\linewidth, height=.11\linewidth]{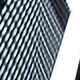}
& \includegraphics[width=.11\linewidth, height=.11\linewidth]{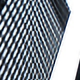}
\\
& QuickSR-M  & ETDS-M & PlainUSR-S  
& \quad 
&   & QuickSR-M  & ETDS-M & PlainUSR-S  
\\
& {\includegraphics[width=.11\linewidth, height=.11\linewidth]{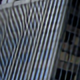}} 
& \includegraphics[width=.11\linewidth, height=.11\linewidth]{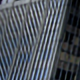}
& \includegraphics[width=.11\linewidth, height=.11\linewidth]{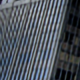}
&   &
& {\includegraphics[width=.11\linewidth, height=.11\linewidth]{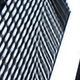}} 
& \includegraphics[width=.11\linewidth, height=.11\linewidth]{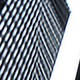}
& \includegraphics[width=.11\linewidth, height=.11\linewidth]{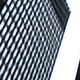}
\\
{\emph{img\_061}~\cite{Urban100}}  
& QuickSR-L & ETDS-L & PlainUSR-M  
&  
& {\emph{img\_062}~\cite{Urban100}}
& QuickSR-L & ETDS-L & PlainUSR-M  
\\
\end{tabular}
\caption{Visual comparison for efficient SR models with an upscaling factor $\times$2. (Best viewed by zooming)}
\label{fig:Bi_fig2}
\end{figure*}

\begin{table*}[!t]
\fontsize{7.0pt}{9.5pt}\selectfont
\tabcolsep=1pt
\begin{center}
\begin{tabular}{l|c|ccc|c|c|c|c}
\whline
\multirow{2}{*}{Method} & Latency & {Para} & {MACs} & Acts & {Set5~\cite{Set5}} &  {Set14~\cite{Set14}} & {BSD100~\cite{B100}} & {Urban100~\cite{Urban100}}
\\
% \cline{5-9}
& (ms) & (K) & (G) & (M) & PSNR/SSIM & PSNR/SSIM & PSNR/SSIM & PSNR/SSIM
\\
\whline
FDIWN~\cite{FDIWN} & 821.6
& 629.2 & 77.58 & {1008.1}  
& 38.07/0.9608
& 33.75/0.9201
& 32.23/0.9003
& 32.40/0.9305
\\
NGswin~\cite{NGSwin} & 1104.9
& 998.4 & 111.09 & -  
& 38.05/\red{0.9610}
& 33.79/0.9199
& 32.27/0.9008
& 32.53/\red{0.9324}
\\
CoMoNet-S~\cite{CoMo} & 831.4
& 722.1 & 58.70 & {273.68}  
& 38.06/0.9606
& \red{33.84/0.9206}
& 32.23/0.9003
& 32.45/0.9315
\\
\rowcolor{tablered}
PlainUSR-L & \red{67.4}
& 707.1 & 169.84 & {78.29}  
& \red{38.07}/0.9607
& {33.82/0.9202}
& \red{32.27/0.9009}
& \red{32.53}/0.9320
\\
\whline
SwinIR-light~\cite{SwinIR} & 3096.9
& 910.2 & 217.51 & -  
& \red{38.14/0.9611}
& 33.86/0.9206
& 32.31/0.9012
& \red{32.76/0.9340}
\\
\rowcolor{tablered}
PlainUSR-L$+$ & \red{257.6}
& 707.1 & 169.84 & {78.29}  
& 38.13/0.9610
& \red{33.88/0.9211}
& \red{32.32/0.9014}
& 32.73/0.9335
\\
\whline
\end{tabular}
% \vspace{2mm}
\end{center}
\caption{Quantitative comparison with quality-oriented models for efficient SR ($\times$2).}
\label{tab:sr3}
\vspace{-4mm}
\end{table*}

\begin{figure*}[!t]
\tabcolsep=1.0pt
\centering
\fontsize{6.0pt}{9.5pt}\selectfont
\renewcommand\arraystretch{0.8}
\begin{tabular}{ccccccccccccc}
\multirow{-5.3}{*}
{\includegraphics[width=.14\linewidth, height=.25\linewidth]{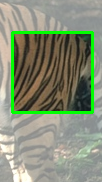}} 
& \includegraphics[width=.11\linewidth, height=.11\linewidth]{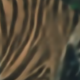}
& \includegraphics[width=.11\linewidth, height=.11\linewidth]{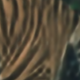}
& \includegraphics[width=.11\linewidth, height=.11\linewidth]{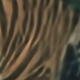}
&
    
&\multirow{-5.3}{*} {\includegraphics[width=.14\linewidth, height=.25\linewidth]{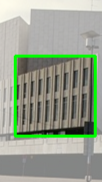}} 
& \includegraphics[width=.11\linewidth, height=.11\linewidth]{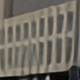} 
& \includegraphics[width=.11\linewidth, height=.11\linewidth]{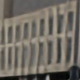}
& \includegraphics[width=.11\linewidth, height=.11\linewidth]{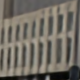}
\\
  & ECBSR-{\tiny{C64}} & DIPNet & PlainUSR-B 
& \quad 
&   & ECBSR-{\tiny{C64}} & DIPNet & PlainUSR-B 
\\
& {\includegraphics[width=.11\linewidth, height=.11\linewidth]{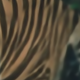}} 
& \includegraphics[width=.11\linewidth, height=.11\linewidth]{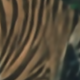}
& \includegraphics[width=.11\linewidth, height=.11\linewidth]{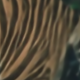}
&   &
& {\includegraphics[width=.11\linewidth, height=.11\linewidth]{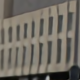}} 
& \includegraphics[width=.11\linewidth, height=.11\linewidth]{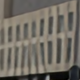}
& \includegraphics[width=.11\linewidth, height=.11\linewidth]{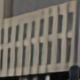}
\\
{\emph{108005}~\cite{B100}}  
& IMDN & DDistill-SR & PlainUSR-L 
&  
& {\emph{78004}~\cite{B100}}
& IMDN & DDistill-SR & PlainUSR-L  
\\
\multirow{-5.3}{*}
{\includegraphics[width=.14\linewidth, height=.25\linewidth]{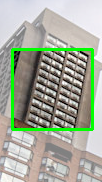}} 
& \includegraphics[width=.11\linewidth, height=.11\linewidth]{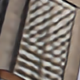}
& \includegraphics[width=.11\linewidth, height=.11\linewidth]{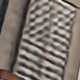}
& \includegraphics[width=.11\linewidth, height=.11\linewidth]{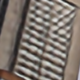}
&
    
&\multirow{-5.3}{*} {\includegraphics[width=.14\linewidth, height=.25\linewidth]{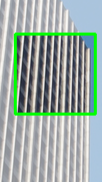}} 
& \includegraphics[width=.11\linewidth, height=.11\linewidth]{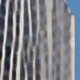} 
& \includegraphics[width=.11\linewidth, height=.11\linewidth]{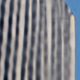}
& \includegraphics[width=.11\linewidth, height=.11\linewidth]{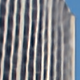}
\\
  & ECBSR-{\tiny{C64}} & DIPNet & PlainUSR-B 
& \quad 
&   & ECBSR-{\tiny{C64}} & DIPNet & PlainUSR-B 
\\
& {\includegraphics[width=.11\linewidth, height=.11\linewidth]{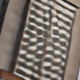}} 
& \includegraphics[width=.11\linewidth, height=.11\linewidth]{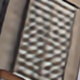}
& \includegraphics[width=.11\linewidth, height=.11\linewidth]{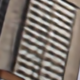}
&   &
& {\includegraphics[width=.11\linewidth, height=.11\linewidth]{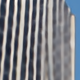}} 
& \includegraphics[width=.11\linewidth, height=.11\linewidth]{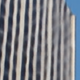}
& \includegraphics[width=.11\linewidth, height=.11\linewidth]{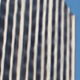}
\\
{\emph{img\_020}~\cite{Urban100}}  
& IMDN & DDistill-SR & PlainUSR-L 
&  
& {\emph{img\_096}~\cite{Urban100}}
& IMDN & DDistill-SR & PlainUSR-L  
\\
\multirow{-5.3}{*}
{\includegraphics[width=.14\linewidth, height=.25\linewidth]{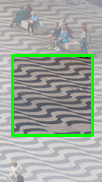}} 
& \includegraphics[width=.11\linewidth, height=.11\linewidth]{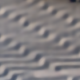}
& \includegraphics[width=.11\linewidth, height=.11\linewidth]{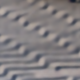}
& \includegraphics[width=.11\linewidth, height=.11\linewidth]{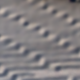}
&
    
&\multirow{-5.3}{*} {\includegraphics[width=.14\linewidth, height=.25\linewidth]{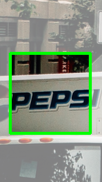}} 
& \includegraphics[width=.11\linewidth, height=.11\linewidth]{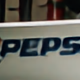} 
& \includegraphics[width=.11\linewidth, height=.11\linewidth]{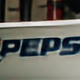}
& \includegraphics[width=.11\linewidth, height=.11\linewidth]{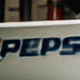}
\\
& ECBSR-{\tiny{C64}} & DIPNet & PlainUSR-B 
& \quad 
&   & ECBSR-{\tiny{C64}} & DIPNet & PlainUSR-B 
\\
& {\includegraphics[width=.11\linewidth, height=.11\linewidth]{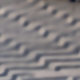}} 
& \includegraphics[width=.11\linewidth, height=.11\linewidth]{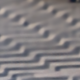}
& \includegraphics[width=.11\linewidth, height=.11\linewidth]{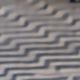}
&   &
& {\includegraphics[width=.11\linewidth, height=.11\linewidth]{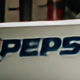}} 
& \includegraphics[width=.11\linewidth, height=.11\linewidth]{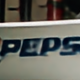}
& \includegraphics[width=.11\linewidth, height=.11\linewidth]{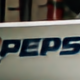}
\\
{\emph{1261}~\cite{div8k}}  
& IMDN & DDistill-SR & PlainUSR-L 
&  
& {\emph{0861}~\cite{div2k}}
& IMDN & DDistill-SR & PlainUSR-L  
\\
\end{tabular}
\caption{Visual comparison for efficient SR models with an upscaling factor $\times$4. (Best viewed by zooming)}
\label{fig:Bi_fig}
\end{figure*}

\noindent\textbf{More comparison with quality-oriented methods}.
In the above comparisons, we compare the proposed PlainUSR with models trading off quality and latency. Here, we provide a detailed comparison with more quality-oriented methods (\eg, transformers and dedicated ConvNet) in~\cref{tab:sr3}. In a word, PlainUSR has the lowest latency and highest quality. Specifically, compared to NGswin~\cite{NGSwin}, the PlainUSR-L is 16.4$\times$ faster and attains higher PSNR/SSIM on BSD100. Compared to SwinIR-light, the PlainUSR-L+ accelerates by 12.0$\times$ and maintains similar restoration quality.

\noindent\textbf{Qualitative comparison}.
In \cref{fig:Bi_fig2,fig:Bi_fig}, we display the visual comparison between the PlainUSR family with other SOTA approaches. Generally, our methods recover substantially clearer and more accurate structural content than others. Concretely, for \emph{barbara} from \cref{fig:Bi_fig2}, PlainUSR-S/M are the only two models restoring the correct cloth texture. For \emph{img\_020} in \cref{fig:Bi_fig}, PlainUSR-B/L recovers more legible patterns of windows than models with similar latency.

\subsection{Ablation studies}
\noindent\textbf{Effectiveness of RepMBConv}.
In \cref{tab:ab_mbconv}, we offer the roadmap of transferring RepMBConv into convolution and more comparisons with other reparameterization strategies like RepVGG~\cite{RepVGG}, ECB~\cite{ECB}, and RRRB~\cite{FMEN} on the proposed PlainUSR. First, we compare the performance between vanilla convolution and MBConv~\cite{mobilenetv3}. While  MBConv leads by over 0.05dB and is lighter than standard convolution, the latter is in turn 3$\times$ faster due to the lower memory access and activation. To endow MBConv with faster inference, \ie, convolution, we replace the nonlinear layers and use the complex module to propose RepMBConv. Replacing nonlinear makes the module reparameterizable but induces a huge performance drop (0.13dB on DIV2KV). Employing complex modules alleviates and even improves the performance decrease. Overall, the proposed RepMBConv can attain similar performance as the original MBConv but maintain the inference as a single convolution. Moreover, the RepMBConv surpasses other RepConv by over 0.06dB/0.002 PSNR/SSIM improvement on DIV2KV, demonstrating its effectiveness.

\begin{table*}[!t]
\center
\small
\footnotesize
%\scriptsize
\fontsize{7.0pt}{9.5pt}\selectfont
\tabcolsep=1.5pt
\begin{center}
\caption{Quantitative comparison between varied parameterization, \ie RepConv \vs RepMBConv on PlainUSR for $\times$4 SR tasks. For ``$\dagger$'' items, the metrics \{Para, MACs, Mem, Acts\} are reported at the training time.}

\label{tab:ab_mbconv}
  % Here 1/2

\begin{tabular}{l|c|cccc|c|c|c}
\whline
\multirow{2}{*}{Method} & Latency & {Para} & {MACs} & {Mem} & Acts & {DIV2KV~\cite{div2k}} &  {Test2K~\cite{div8k}} & {Test4K~\cite{div8k}}
\\
% \cline{5-9}
& (ms) & (K) & (G) & (M) & (M) & PSNR/SSIM & PSNR/SSIM & PSNR/SSIM 
\\
\whline
Convolution & 26.8
& 333 & 18.69 & {327.7} & {46.93}  
& 28.81/0.8148
& 26.08/0.7550
& 27.51/0.8021
\\
RepVGG$^\dagger$~\cite{RepVGG} & 26.8 
& 583 & 35.00 & {328.7} & {80.48}  
& 28.80/0.8147
& 26.09/0.7554
& 27.53/0.8025
\\
ECB$^\dagger$~\cite{ECB} & 26.8 
& 989 & 61.27 & {479.2} & 348.9  
& 28.83/0.8153
& 26.10/0.7557
& 27.54/0.8028
\\
RRRB$^\dagger$~\cite{FMEN} & 26.8 
& 1192 & 74.86 & {649.1} & {181.1}  
& 28.83/0.8151
& 26.10/0.7558
& 27.54/0.8030
\\
\whline
% \hline
MBConv~\cite{mobilenetv3} & 78.3 %& 32.6 %& 
& 261 & 10.24 & 399.1 & 181.1
& 28.88/0.8166 %30.59/0.8409
% & 26.29/0.7141 %27.63/0.7381
& 26.12/0.7568 %27.65/0.7762
& 27.57/0.8039
%\hline
\\
{Replacing Nonlinear}$^\dagger$ & 26.8 %& 32.6 %& 
& 263 & 10.24 & 399.1 & 181.1
& 28.75/0.8133 %30.59/0.8409
% & 26.29/0.7141 %27.63/0.7381
& 26.06/0.7539 %27.65/0.7762
& 27.49/0.8014
%\hline
\\
{Complex Module}$^\dagger$ & 26.8 %& 32.6 %& 
& 1249 & 74.86 & 658.4 & 181.1
& 28.89/0.8171
& 26.13/0.7573
& 27.58/0.8043
%\hline
\\
% \rowcolor{tablered}
% RepMBConv (Training) & 26.8 %& 32.6 %& 
% & 333 & 18.69 & {327.7} & {46.93}  
% & 28.96/0.8181 %30.42/0.8373
% % & 26.20/0.7111 %27.54/0.7351
% & 26.18/0.7579 %27.59/0.7728
% % & 30.24/0.9038
% & 27.63/0.8048
% %\hline
% \\
\rowcolor{tablered}
RepMBConv & 26.8 %& 32.6 %& 
& 333 & 18.69 & {327.7} & {46.93}  
& 28.89/0.8171
& 26.13/0.7573
& 27.58/0.8043
% & 28.96/0.8181 %30.42/0.8373
% % & 26.20/0.7111 %27.54/0.7351
% & 26.18/0.7579 %27.59/0.7728
% % & 30.24/0.9038
% & 27.63/0.8048
\\
\whline 
\end{tabular}
\end{center}
\end{table*}

\begin{table*}[!t]
\center
\small
\footnotesize
%\scriptsize
\fontsize{7.0pt}{9.5pt}\selectfont
\tabcolsep=1pt
\begin{center}
\caption{Quantitative comparison between varied attention modules on $\times$4 SR tasks.}

\label{tab:ab_lip}
  % Here 1/2

\begin{tabular}{l|c|c|cccc|c|c|c}
\whline
\multirow{2}{*}{Method} & \multirow{2}{*}{Attention} & Latency & {Para} & {MACs} & {Mem} & Acts & {DIV2KV~\cite{div2k}} &  {Test2K~\cite{div8k}} & {Test4K~\cite{div8k}}
\\
% \cline{5-9}
&  & (ms) & (K) & (G) & (M) & (M) & PSNR/SSIM & PSNR/SSIM & PSNR/SSIM 
\\
\whline
\multirow{6}{*}{PlainUSR} & Baseline & 20.2 %& 32.6 %& 
& 280 & 18.34 & 262.5 & 41.09
& 28.61/0.8092 
& 26.00/0.7508 
& 27.38/0.7978
%\hline
\\
& SE~\cite{SE} & 22.5 %& 32.6 %& 
& 287 & 18.34 & 270.7 & 41.09
& 28.62/0.8095 %30.59/0.8409
% & 26.29/0.7141 %27.63/0.7381
& 26.00/0.7509 %27.65/0.7762
& 27.38/0.7978
%\hline
\\
& CBAM~\cite{CBAM} & 55.7 %& 32.6 %& 
& 289 & 18.37 & 263.6 & 41.42
& 28.64/0.8097 %30.59/0.8409
% & 26.29/0.7141 %27.63/0.7381
& 25.99/0.7506 %27.65/0.7762
& 27.36/0.7972
%\hline
\\
& ESA~\cite{CBAM} & 30.9 %& 32.6 %& 
& 337 & 19.20 & 395.5 & 70.05
& 28.86/0.8157 %30.59/0.8409
% & 26.29/0.7141 %27.63/0.7381
& 26.11/0.7562 %27.65/0.7762
& 27.56/0.8032
%\hline
\\
& NLSA~\cite{NLSA} & 631.5
& 326 & 21.28 & 6748.9 & 62.06
& 28.91/0.8173 
& 26.13/0.7575
& 27.59/0.8043
\\
\rowcolor{tablered}
\cellcolor{white} & LIA & 26.8 
& 333 & 18.69 & {327.7} & {46.93}  
& 28.89/0.8171
& 26.13/0.7573
& 27.58/0.8043
\\
\whline
& ESA~\cite{RFANet}  & 56.5 
& 433 & 27.10 & 832.0 & 112.03
& 28.96/0.8182 
& 26.17/0.7585 
& 27.64/0.8054
\\
\rowcolor{tablered}
\multirow{-2}{*}{RFDN~\cite{RFDN}} 
\cellcolor{white}
& LIA & 50.3 
& 449 & 26.92 & {755.3} & {96.28}  
& 28.97/0.8185 
& 26.19/0.7588 
& 27.65/0.8057
%\hline
\\
\whline
\end{tabular}
\end{center}
\end{table*}

\begin{figure}[!t]
\centering
\scriptsize
\tabcolsep = 1pt
\begin{tabular}{cccc}
\includegraphics[height=0.172\linewidth]{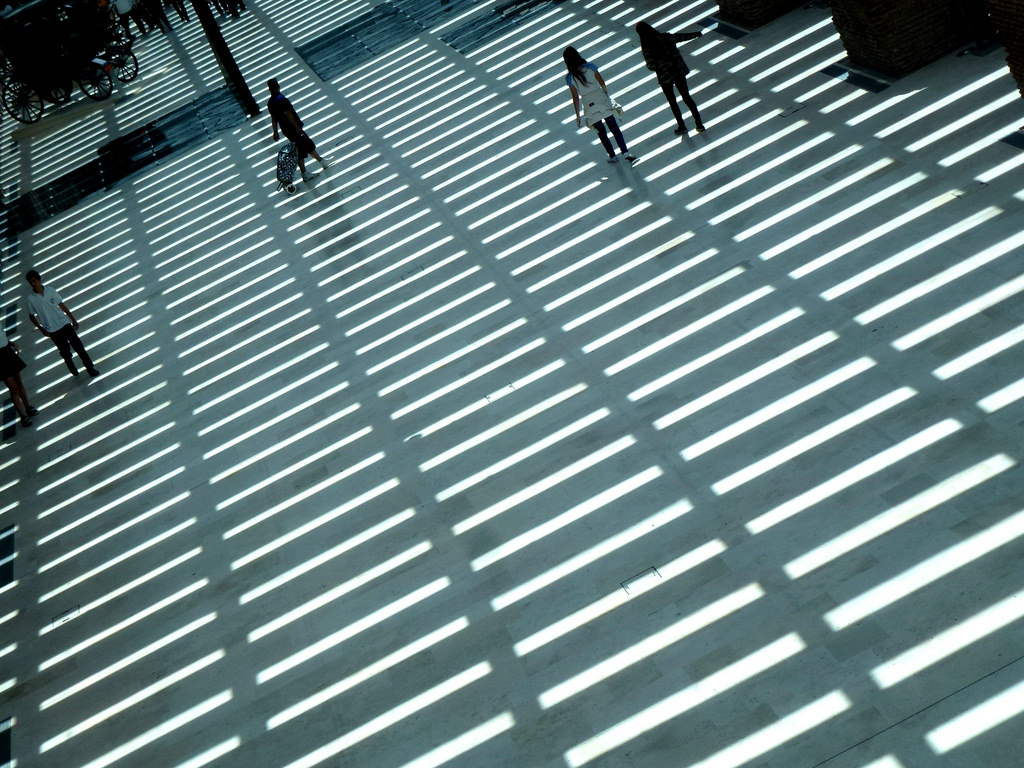}
& \includegraphics[height=0.172\linewidth]{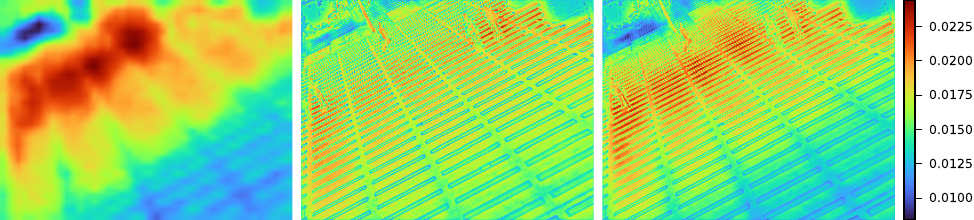}
\\
LR & $\psi(\sigma(\mathcal{I(\rmX)}))$\quad\quad\quad\quad\quad\quad$\sigma(\rmX_{[0]})$\quad\quad\quad\quad$ \psi(\sigma(\mathcal{I(\rmX)}))\odot\sigma(\rmX_{[0]})$\\
\end{tabular}
\caption{Visual activation maps of each term in~\cref{eq:lia}.}
\label{fig:ab_lia}
\end{figure}

\noindent\textbf{Effectiveness of LIA}. To investigate the effectiveness of LIA, we compare it with several existing attention modules, including SE~\cite{SE}, CBAM~\cite{CBAM}, ESA~\cite{RFANet}, and NLSA~\cite{NLSA} on the proposed PlainUSR and RFDN~\cite{RFDN} in \cref{tab:ab_lip}. In general, the LIA can catch up with dedicated 2-order attention and surpasses 1-oder attention by large margins. Particularly, the PSNR of LIA decreases a maximum of 0.02dB than NLSA while accelerating 23$\times$. Compared to ESA, the proposed LIA spends fewer calculations and time and obtains notable gains over three challenging benchmarks on both PlainUSR and RFDN. These numerical results imply the efficiency and effectiveness of the LIA module.

To evaluate the influence of individual components of LIA, we ablate them in \cref{tab:ab_lia} with six variants. The variant \Rmnum{1} is the baseline without attention mechanism. For variant \Rmnum{2} and \Rmnum{3}, we degrade LIA with only 1-order information interaction. Removing any of them (local importance or gate mechanism) will result in huge performance drops. Moreover, local importance plays a more important role in restoration. For variant \Rmnum{4}, we replace the softmax-based importance with the max-based one, inducing a 0.1dB/0.004 decline on Urban100. For variant \Rmnum{5}, we swap the sigmoid activation and bilinear interpolation. This operation has no impact on performance but increases the 1.7ms latency. In~\cref{fig:ab_lia}, we exhibit the activation maps for \cref{eq:lia}, where the local importance captures the approximate responses and then gate rectifies it for more precise attention.

\noindent\textbf{Effectiveness of PlainU-Net backbone}. In \cref{tab:backbone,fig:training}, we present the comparison between the proposed PlainU-Net with the other two style backbones: VGG~\cite{RepVGG}-style and ResNet~\cite{ResNet}-style. We set the channel number 32 for these two backbone and \{48,32,16\} for PlainU-Net to maintain similar complexity. After fair training under the same settings, we observe that PlainU-Net provides a better-optimizing curve than the other two backbones. The PSNR improvements of PlainU-Net over VGG/ResNet are 0.09dB/0.08dB on Urban100.

\begin{table*}[!t]
\center
\small
\footnotesize
%\scriptsize
\fontsize{7.0pt}{9.5pt}\selectfont
\tabcolsep=1.5pt
\begin{center}
\caption{Quantitative comparison between varied LIA variants on $\times$4 SR tasks.}

\label{tab:ab_lia}
  % Here 1/2

\begin{tabular}{c|c|c|cccc|c|c}
\whline
\multirow{2}{*}{Variant} & \multirow{2}{*}{$\mathcal{A}(\rmX)$} & Latency & {Para} & {MACs} & {Mem} & Acts & {Urban100~\cite{Urban100}} &  {Test2K~\cite{div8k}}
\\
% \cline{5-9}
& & (ms) & (K) & (G) & (M) & (M) & PSNR/SSIM & PSNR/SSIM 
\\
\whline
\Rmnum{1} & $\rmI\odot\rmX$ & 20.2 %& 32.6 %& 
& 280 & 18.34 & 262.5 & 41.09
& 25.43/0.7634 
& 26.00/0.7508 
% & 27.38/0.7978
%\hline
\\
\Rmnum{2} & $\sigma(\rmX_{[0]}) \odot \rmX$ & 21.9 %& 32.6 %& 
& 280 & 18.34 & 262.5 & 41.09
& 25.54/0.7671 
& 26.03/0.7522 
% & 27.38/0.7978
%\hline
\\
\Rmnum{3} & $\psi(\sigma(\mathcal{I(\rmX)})) \odot \rmX$ & 24.6
& 333 & 18.69 & 262.7 & 46.93
& 25.68/0.7727
& 26.07/0.7543
\\
\Rmnum{4} & $\sigma(\rmX_{[0]}) \odot \psi(\sigma(\mathcal{\red{I'}(\rmX)})) \odot \rmX$ & 25.9 
& 333 & 18.69 & {327.7} & {46.93}  
& 25.78/0.7760
& 26.09/0.7552
\\
\Rmnum{5} & $\sigma(\rmX_{[0]}) \odot \red{\sigma}(\red{\psi}(\mathcal{I(\rmX)})) \odot \rmX$ & 28.5 %& 32.6 %& 
& 333 & 18.69 & {327.7} & {46.93}  
& 25.88/0.7800
& 26.13/0.7573
\\
\rowcolor{tablered}
\Rmnum{6} & $\sigma(\rmX_{[0]}) \odot \psi(\sigma(\mathcal{I(\rmX)})) \odot \rmX$ & 26.8 
& 333 & 18.69 & {327.7} & {46.93}  
& 25.88/0.7800
& 26.13/0.7573
\\
\whline
\end{tabular}
\end{center}
\end{table*}

\begin{table}[t]
\begin{minipage}[c]{0.66\textwidth}
%\scriptsize
\fontsize{7.0pt}{9.5pt}\selectfont
\tabcolsep=1.2pt
  % Here 1/2

\begin{tabular}{l|c|cc|c|c}
\whline
\multirow{2}{*}{Variant} & Latency & {Mem} & Acts & {Urban100~\cite{Urban100}} &  {Test2K~\cite{div8k}}
\\
% \cline{5-9}
& (ms) & (M) & (M) & PSNR/SSIM & PSNR/SSIM 
\\
\whline
VGG~\cite{RepVGG}-style & 17.7 %& 32.6 %& 
& 167.2 & 29.39
& 25.51/0.7672 
& 26.00/0.7519 
% & 27.38/0.7978
%\hline
\\
ResNet~\cite{EDSR}-style & 19.8 %& 32.6 %& 
& 228.5 & 29.19
& 25.52/0.7674 
& 26.02/0.7523 
% & 27.38/0.7978
%\hline
\\
\rowcolor{tablered}
PlainU-Net & 18.1 
& {247.2} & {32.56}  
& 25.60/0.7702
& 26.04/0.7531
\\
\whline
\end{tabular}
\vspace{2mm}
\captionof{table}{Quantitative comparison between varied backbones on $\times$4 SR tasks.}
\label{tab:backbone}
\end{minipage}
\quad
\begin{minipage}[r]{0.3\textwidth}
\includegraphics[width=0.95\linewidth]{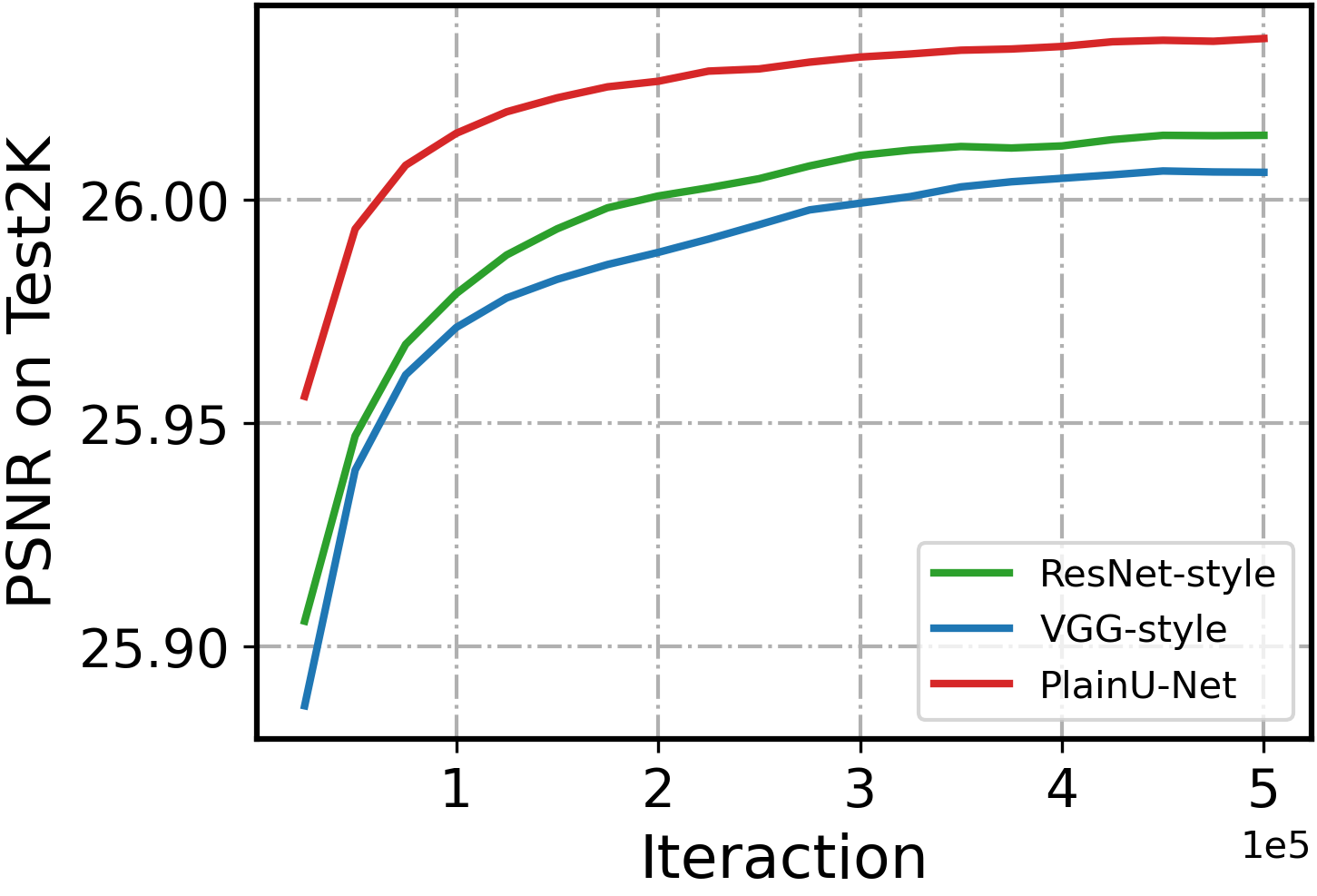}
\captionof{figure}{Training curve.}
\label{fig:training}
\end{minipage}
\vspace{-6mm}
\end{table}

\section{Conclusion}
This paper presents PlainUSR, a novel framework chasing faster ConvNets for image super-resolution tasks. Our methodology focuses on reducing inference latency through three key modifications: RepMBConv, LIA module, and PlainU-Net, targeting convolution, attention, and backbone components. Through comprehensive experiments, we demonstrate that each of these modifications surpasses existing schemes in both inference latency and reconstruction quality. By synergistically integrating them, the PlainUSR family achieves remarkable trade-offs between latency, complexity, and quality.
% This underscores the efficacy of PlainUSR as a promising advancement in the realm of faster and more efficient image super-resolution ConvNets.

 % TODO REVIEW/FINAL: This \clearpage needs to be removed from both review and camera-ready versions.

\bibliographystyle{splncs04}
\bibliography{egbib}
\end{document}